\documentclass[a4paper]{jpconf}

\usepackage{amsmath,graphicx,cite,bbold}
\usepackage{color}

\begin{document}
\title{$S_3$ as a unified family theory for quarks and leptons}
\author{F Gonz\'alez Canales$^{1,2}$, A Mondrag\'on$^{1}$, U J Salda\~na Salazar$^{1}$,\\ L Velasco-Sevilla$^{3}$}
\address{{(1) Instituto de F\'{\i}sica, Universidad Nacional Aut\'onoma de M\'exico,}\\
{ Apdo. Postal 20-364, 01000, M\'exico D.F., M\'exico.}\\
{(2) Facultad de Ciencias de la Electr\'onica, Benem\'erita Universidad Aut\'onoma de Puebla,}\\
{Apdo. Postal 157, 72570, Puebla, Pue., M\'exico.}\\
{(3) CINVESTAV-IPN, Apdo.~Postal 14-740, 07000, M\'exico D.F., M\'exico.}}

\ead{ulisesjesus@fisica.unam.mx}

	\begin{abstract}
			We  present an $S_3$-invariant extension of the SM which is able to account for the mixing both in the quark and the lepton sector. We focus here on the quark sector and present different realizations of the model, according to how many Higgs fields are involved. We  then confront the models with the up-to-date values of masses and mixing in the quark sector.
\end{abstract}

	\section{Introduction}
	The long awaited remaining piece of the Standard Model (SM), \textit{i.e.} the Higgs particle, has been probably recently detected \cite{Chatrchyan,Aad}. Despite this great discovery, the SM, as was originally formulated,  needs to be extended in order to accommodate the small masses of neutrinos. Moreover, clearly the number of free parameters needed to account for the values of fermion masses and mixing parameters ($12$ masses, $6$ mixing angles, and $2$ ($4$) Dirac phases (Dirac plus Majorana phases)), reflects our lack of understanding of the flavour puzzle.  Nevertheless, at present, mathematical relations among masses and mixing angles \cite{Fritzsch:1999ee,Ishimori:2010au,Altarelli:2010gt,Hirsch:2012ym} seem to indicate a common origin for the pattern of mixing and the hierarchical spectrum of masses. This origin could be justified by introducing additional symmetries into the theory, which are commonly called \textit{horizontal or family symmetries} because they act in the generation or family space of fermions.  
	
	The main goal of this work is to present an extension of the SM 
	where successful  and exact mathematical relations among fermion masses and mixing angles can be obtained. In section 2, we give some brief arguments on how easy the $S_3$ symmetry can be interpreted as a family symmetry.  In section 3, we discuss and introduce the construction of the $S_3$-invariant extension of the Standard Model ($S_3$-SM) we are considering. In section 4, we contrast the different realizations of this model with the most up to date experimental information. In section 5, we summarize and conclude with some remarks. 	 

	\section{$S_3$ as a family symmetry}
			In our universe, atoms and atomic processes, as beta decay, require a single \textit{family} of fundamental fermions ($e$, $\nu_e$, $d$, $u$). However, in subatomic processes nature seems to need two more families of fundamental particles, replicas of the first family but with different masses. The couplings of the three families to the force carriers is universal,  making no distinction between particles with the same quantum charges but which belong to a different family. The SM Lagrangian, except for the terms describing the Yukawa interactions, is invariant under the permutation of the family index, that is, any permutation of the family index should leave the system invariant. The latter statement can be treated as having applied the following symmetry transformations, 
\begin{footnotesize}
		\begin{eqnarray} \label{S3_mtxelements}
			\begin{pmatrix}
				1 & 0 & 0 \\
				0 & 1 & 0 \\
				0 & 0 & 1
			\end{pmatrix} \quad
			\begin{pmatrix}
				0 & 1 & 0 \\
				1 & 0 & 0 \\
				0 & 0 & 1
			\end{pmatrix} \quad
			\begin{pmatrix}
				0 & 0 & 1 \\
				0 & 1 & 0 \\
				1 & 0 & 0
			\end{pmatrix} \quad			
			\begin{pmatrix}
				1 & 0 & 0 \\
				0 & 0 & 1 \\
				0 & 1 & 0
			\end{pmatrix} \quad
			\begin{pmatrix}
				0 & 0 & 1 \\
				1 & 0 & 0 \\
				0 & 1 & 0
			\end{pmatrix} \quad
			\begin{pmatrix}
				0 & 1 & 0 \\
				0 & 0 & 1 \\
				1 & 0 & 0
			\end{pmatrix},			
		\end{eqnarray}
\end{footnotesize} 

\noindent which leave invariant an equilateral triangle whose vertices are the unitary basis vectors. This last statement expresses what we call an isomorphism between the permutation group of three objects, $S_3$, and the equilateral triangle symmetry group, $D_3$. We will later refer  to this basis, where each family is undistinguishable, as the \textit{universal-basis}. 

			\subsection{$S_3$ theoretical aspects}
			
					$S_3$ is the smallest non-abelian discrete symmetry group. It has $3! = 6$ elements and has two classes of non-trivial transformations: even and odd permutations, $A$ and $B$, respectively. The two kind of generators, $A$ and $B$, fulfill the relations
		\begin{equation}
			A^2 = B^3 = 1, \quad\quad A B A = B^{-1}.
		\end{equation}	
		When taking into account the trivial transformation, it is found that $S_3$ contains three classes of transformations; thus, it has three irreducible representations (irreps): a doublet $\bf{2}$, and two singlets, ${\bf{1}}_S$ and ${\bf{1}}_A$, symmetric and antisymmetric, respectively. 
		
		The Kronecker products between irreps may be decomposed as a direct sum of irreps: 
		\begin{eqnarray}\nonumber		
		{\bf{1}}_{S} \otimes {\bf{1}}_{S} = {\bf{1}}_{S}, \quad {\bf{1}}_{A} \otimes {\bf{1}}_{A}= {\bf{1}}_{S}, \quad {\bf{1}}_{A} \otimes {\bf{1}}_{S}= {\bf{1}}_{A},\hspace{.7cm}\\
		 {\bf{1}}_{S} \otimes {\bf{2}} = {\bf{2}}, \quad {\bf{1}}_{A} \otimes {\bf{2}}= {\bf{2}}, \quad \text{and} \quad {\bf{2}} \otimes {\bf{2}} =  {\bf{1}}_{A} \oplus {\bf{1}}_{S} \oplus {\bf{2}}.
		\end{eqnarray}

			\subsection{The flavour-basis}
			Fermions have different masses, in fact, so different that are commonly referred as hierarchical masses,
\begin{footnotesize}
			\begin{equation}
				\begin{pmatrix}
					m_e \\
					m_{\nu_1} \\
					m_d \\
					m_u \\
				\end{pmatrix} \quad << \quad
				\begin{pmatrix}
					m_\mu \\
					m_{\nu_2} \\
					m_s \\
					m_c \\
				\end{pmatrix} \quad << \quad
				\begin{pmatrix}
					m_\tau \\
					m_{\nu_3} \\
					m_b \\
					m_t \\
				\end{pmatrix},
			\end{equation}
\end{footnotesize}

		\noindent where we are assuming, just for illustration, a normal hierarchy and non degenerate neutrino masses. After the introduction of the Yukawa interactions, each fermion should couple differently to the Higgs field, such that, when the electroweak symmetry gets spontaneuosly broken, each fermion gains a different mass. 
	A unitarity transformation takes us from the \textit{universal-basis} to the \textit{$S_3$ symmetry adapted basis}, where all the elements in Equation~\ref{S3_mtxelements} appear now as a direct sum of irreducible representations, ${\bf 2} \oplus 1_{\bf S}$. It is in this new basis, where the interaction eigenstates are defined, the new eigenstates may also be called  flavour eigenstates (in regard to flavour oscillations). Hereafter, we will refer to this basis as the \textit{flavour basis}. 
			
			The interaction eigenstates in terms of the universal eigenstates are defined as
			\begin{eqnarray} F_{\bf D} =
				\begin{pmatrix}
					F_{1} \\
					F_{2}
				\end{pmatrix} =
				\begin{pmatrix}
					\frac{1}{\sqrt{2}}(f_1 - f_2) \\
					\frac{1}{\sqrt{6}}(f_1 + f_2 - 2f_3)
				\end{pmatrix} \quad \text{and} \quad
				F_{\bf S} = \frac{1}{\sqrt{3}}(f_1 + f_2 + f_3).
			\end{eqnarray}
			 Note that we have not yet defined which flavour eigenstate corresponds to a certain family. We will  make this choice later.
			 		
	\section{An $S_3$-invariant extension of the Standard Model}
	 Models with $S_3$ as their family symmetry have been widely studied, see references therein~\cite{Hirsch:2012ym}. 
Our approach is somewhat similar to the one used in~\cite{Yahalom:1983kf}, where different models generated by a different number of Higgs $SU(2)_L$ doublets were explored. Nevertheless, we consider a different assignment of fermions to the irreducible representations of $S_3$. Here we will show that a weak-basis transformation may help us to reduce the number of free parameters, such that, mixing angles can be expressed in terms of mass ratios. We will  focus on the general case of adding three more Higgs $SU(2)_L$ doublets aside from the SM one, briefly mentioning how it compares to the case of the addition of two Higgs doublets \footnote{On a separate work we will present a study of all the possible cases with different number of Higgs $SU(2)_L$ doublets.}. The four Higgs fields are accommodated such that they occupy all the irreps of $S_3$: $H_{DW} = (H_{1W}, H_{2W})^T$, $H_{SW}$, and $H_{AW}$, where the subindexes $D$, $S$, and $A$ refer respectively to the doublet, symmetric and antisymmetric $S_3$ representations, and the subindex $W$ means that the field behaves as a doublet of $SU(2)_L$.	

 

		\subsection{Assignments between $S_3$ irreps and families}
		We need to decide how the assignment between the three families with the $S_3$ irreps should be done. {For it, we recall the following fact: for each kind of fermion, there exist three identical particles, except for their mass.} 
	{	In fact, if we divide the mass of each quark over the mass of the heaviest quark of each kind, 
		we can detect a clear hierarchy among their masses: the first two families are in fact really close to each other when compared to the third one.} 
%
		This suggests that the fermion masses of each type could be reflecting an $S_3$ symmetric pattern: ${\bf 2} \oplus {\bf 1}$. 
		
		Hence, we generically assign the first two families, $f_{I(L,R)}$ and $f_{II(L,R)}$, to the doublet 
 representation, $\bf{2}$, and the third family, $f_{III(L,R)} \equiv g_{L,R}$, to the symmetric singlet 
 representation, ${\bf{1}}_{S}$,
 \begin{equation}
  \begin{pmatrix}
	f_{I(L,R)} \\
	f_{II(L,R)}
  \end{pmatrix} \sim {\bf 2}; \quad 
  f_{III(L,R)} \equiv g_{L,R} \sim 1_{\bf S}, \nonumber
 \end{equation}
 respectively. Here $I, II$ or  $III$ represent the family index of a left or right-handed fermion field 
 $f_{(L,R)}$, and $f_{(L,R)}$ may represent any SM fermion field. 
  
		\subsection{A weak-basis transformation: the hierarchichal-basis}
		 Fermions of the SM become massive after the Higgs mechanism. The implied generic mass matrix form for both quarks and leptons is
\begin{footnotesize}
  			\begin{equation}
  {\mathcal{M}}_{S_3}^{f} = \begin{pmatrix}
  \sqrt{2}{Y}_{2}^{f} v_S + {Y}_{3}^{f} w_2 & {Y}_{3}^{f} w_1 + \sqrt{2} {Y}_{4}^{f} v_A & 
  \sqrt{2} {Y}_{5}^{f} w_1 \\
  {Y}_{3}^{f} w_1 - \sqrt{2} {Y}_{4}^{f} v_A & \sqrt{2} {Y}_{2}^{f} v_S -  {Y}_{3}^{f} w_2 & 
  \sqrt{2} {Y}_{5}^{f} w_2 \\
  \sqrt{2} {Y}_{6}^{f} w_1 & \sqrt{2} {Y}_{6}^{f} w_2 & 2 {Y}_{1}^{f} v_S
						 \end{pmatrix},
 			\end{equation}
\end{footnotesize}	
\vspace{-.3cm}

\noindent where the ${Y}_{I}^{f}$ are complex Yukawa couplings and $w_1$, $w_2$, $v_S$, and $v_A$ are the vacuum expectation values of the $S_3$ doublet components, the $S_3$ symmetric singlet, and the $S_3$ antisymmetric singlet, respectively. 
We remind the reader the main goal of this work, that is, to confront the up-to-date values of quark masses and mixing to the expressions that relate mixing angles to mass ratios. These relations are obtained by using the three invariants of a mass matrix. However, the abundance of free parameters makes the calculation cumbersome. It is therefore convenient to find first an equivalent form which could simplify  the problem. 

		The weak charged current is the only manifestation in the SM of the interaction of two particles of the same sector (quark or leptonic) but with different electric charge. 
We call weak-basis transformations  (WBT) the similarity transformations of mass matrices that leave invariant the weak charged currents.  It has been already noticed that a suitable WBT may help to reduce the number of free parameters of a mass matrix by introducing a set of zeroes in its entries \cite{Branco:1999nb,Branco:2007nn}. In our case, we find that a WBT in the $S_3$ doublet space could exhibit one of the two desired null matrix elements that reproduce the 4-zero Fritzsch-like texture~\cite{Weinberg:1977hb,Fritzsch:1977za,Fritzsch:1977vd}. This kind of mass matrices are considered to be, along with the Nearest-Neighbour Interaction mass matrix form, one of the two kind of mass matrices that allow a similar treatment of quarks and leptons such that they have a viable phenomenology (\cite{Gupta:2011zzg} and \cite{Babu:2004tn,Simoes:2011zz}), which we call here {\it unified}. 
A further reduction on the number of free parameters may be achieved by demanding Hermiticity to the mass matrix, which  is always possible within any extension of the SM that does not introduce flavour changing right-handed currents~\cite{Fritzsch:1999ee}. 
%
For these reasons, Hermicity is often invoked. Here, we then consider the following  generic Hermitian mass matrix for Dirac fermions:
\begin{footnotesize}
		\begin{equation}
				\label{eq:2tzDirac}
				{\mathcal{M}}_{Hier.}^{f} = \begin{pmatrix}
	 |\mu_1^{f}|+|\mu_2^{f}|{\cos}^2\theta(1-3{\tan}^2\theta) & |\mu_2^{f}|{\text {sin}}\theta {\cos}\theta(3-{\tan}^2\theta) + i|\mu_5^{f}| & 0 \\
	|\mu_2^{f}|{\text {sin}}\theta {\cos}\theta(3-{\tan}^2\theta) - i|\mu_5^{f}| & |\mu_1^{f}|-|\mu_2^{f}|{\cos}^2\theta(1-3{\tan}^2\theta) & \mu_8^{f}{{\sec}\theta} \\
		0 & \mu_8^{f*}{{\sec}\theta} & |\mu_3^{f}|
				\end{pmatrix},
		  \end{equation}
\end{footnotesize}		
		
		\noindent where we have denoted $\mu_1^{f} \equiv \sqrt{2}{Y}_{2}^{f} v_S, \hspace{.1cm} \mu_2^{f} \equiv {Y}_{3}^{f} w_2, \hspace{.1cm} \mu_3^{f} \equiv 2 {Y}_{1}^{f} v_S, \hspace{.1cm}  \mu_5^{f} \equiv \sqrt{2} {Y}_{4}^{f} v_A, $ 	
		$\text{and} \hspace{.1cm} \mu_8^{f} \equiv \sqrt{2}{Y}_{6}^{f} w_1$. Hermiticity implies a particular phase value for the complex Yukawa couplings, while the nullity of the matrix elements, $(1,3)$ and $(3,1)$, is controlled by the relation $\tan\theta = {w_1}/{w_2}$. 
		
		In order to reproduce the 4-zero Fritzsch-like texture we need a zero $(1,1)$ entry. When a unified treatment is pursued, by using the same mass matrix for all fermions, there is not a possible combination of WBTs that could allow having at the same time zero $(1,3)$, $(3,1)$ and $(1,1)$ entries \cite{Branco:2007nn}. Nevertheless, here we show that the origin of the null entry $(1,1)$ could be in fact much simpler. For simplicity, consider a Hermitian $n$-dimensional matrix ${\cal A}_{n \times n}$ which is diagonalized by the unitary matrix ${\cal U}$, 
\begin{footnotesize}
\begin{eqnarray}
	{\cal U} {\cal A}_{n \times n} {\cal U}^{-1} = diag(a_1, a_2,..., a_n),
\end{eqnarray}
\end{footnotesize}\vspace{-.5cm}

\noindent we may see that the unitary matrix diagonalizing ${\cal A}_{n \times n}$ still diagonalizes the resulting matrix $\overline{{\cal A}}_{n \times n}$ after the extraction of any of its diagonal elements,
\begin{footnotesize}
\begin{eqnarray}
	{\cal A}_{ii}{\mathbb{1}_{n \times n}} + {\cal U} \overline{{\cal A}}_{n \times n} {\cal U}^{-1} = diag(a_1, a_2,..., a_n), \\
	{\cal U} \overline{{\cal A}}_{n \times n} {\cal U}^{-1} = diag(\overline{a}_1, \overline{a}_2,..., \overline{a}_n),
\end{eqnarray}
\end{footnotesize}
\vspace{-.5cm}

\noindent where we have denoted $\overline{a}_j \equiv a_j - {\cal A}_{ii}$. This shows how the unitary matrix ${\cal U}$ is independent of the removed entry ${\cal A}_{ii}$. Therefore, by diagonalizing  ${\cal A}_{n \times n}$ or $\overline{{\cal A}}_{n \times n}$ we arrive to the same unitary matrix, but now with an important difference: the new set of eigenvalues is shifted. Hence, we will shift the set of eigenvalues to produce the $(1,1)$ entry equal to zero. This basis, where the mass matrix with the 4-zero Fritzsch-like texture appears is usually called the {\it hierarchical-basis}.
	
		For these proceedings, we focus on the quark sector of the models. When reparameterizing the mass matrix, the produced shift $\Delta_f$, in the set of eigenmasses, is in fact much smaller than one. This allows us to consider the "shifted" approach as a generalization to the standard approach, where no shift is made but where the zero entries  are imposed by new conditions. 
For the purpose of illustrating the relations among mixing parameters and mass ratios, we will consider that the set of shifted eigenmasses is equal to the no-shifted ones,\textit{ i.e.} $\Delta_f = 0$. The complete analysis of the model, $\Delta_f \neq 0$, will be presented  in a separate work. The main success of this kind of  models is that they allow us to express the elements of the quark mixing matrix $V_{CKM}^{th}$ as a explicit function of the quark mass ratios	
 \begin{footnotesize}
\begin{equation}\label{elem:ckm_S3SM}
  \begin{array}{l}
    V_{ ud }^{ ^{th} } = 
     \sqrt{ \frac{ \widetilde{m}_{c} \widetilde{m}_{s} \xi_{1}^u  \xi_{1}^d }{ 
      {\cal D}_{ 1u } {\cal D}_{ 1d } } } 
      + \sqrt{ \frac{ \widetilde{m}_{u} \widetilde{m}_{d} }{ 
      {\cal D}_{ 1u } {\cal D}_{ 1d } } } \left( \sqrt{ \left( 1 - \delta_{ u } \right) 
      \left( 1 - \delta_{d} \right) \xi_{ 1 }^u \xi_{ 1 }^d } + \sqrt{ \delta_{u} \delta_{d} \xi_{ 2 }^u 
      \xi_{ 2 }^d }e^{ i \phi_2 } \right) e^{ i \phi_1 }, \\\\
    V_{us}^{ ^{th} } = 
     - \sqrt{ \frac{ \widetilde{m}_{c} \widetilde{m}_{d} \xi_{ 1 }^u \xi_{ 2 }^d }{ 
      {\cal D}_{ 1u } {\cal D}_{ 2d } } } + \sqrt{ \frac{ \widetilde{m}_{u} \widetilde{m}_{s} }{ 
      {\cal D}_{ 1u } {\cal D}_{ 2d } } } \left( \sqrt{ \left( 1 - \delta_{u} \right) \left( 1 - 
      \delta_{d} \right) \xi_{ 1 }^u \xi_{ 2 }^d} + \sqrt{ \delta_{u} \delta_{d} \xi_{ 2 }^u \xi_{ 1 }^d }e^{ i \phi_2 } 
      \right) e^{ i \phi_1 }, \\\\
    V_{ub}^{ ^{th} } = 
     \sqrt{ \frac{ \widetilde{m}_{c} \widetilde{m}_{d} \widetilde{m}_{s} \delta_{d} \xi_{ 1 }^u }{ 
      {\cal D}_{ 1u } {\cal D}_{ 3d } } } + \sqrt{ \frac{ \widetilde{m}_{u} }{ 
      {\cal D}_{ 1u } {\cal D}_{ 3d } } } \left( \sqrt{ \left( 1 - \delta_{u} \right) \left( 1 - 
      \delta_{d} \right) \delta_{d} \xi_{ 1 }^u } - \sqrt{ \delta_{u} \xi_{ 2 }^u \xi_{ 1 }^d \xi_{ 2 }^d }e^{ i \phi_2 } 
      \right) e^{ i \phi_1 },\\\\
    V_{cd}^{ ^{th} } = 
     - \sqrt{ \frac{ \widetilde{m}_{u} \widetilde{m}_{s} \xi_{ 2 }^u \xi_{ 1}^d }{ 
      {\cal D}_{ 2u } {\cal D}_{ 1d } } } + \sqrt{ \frac{ \widetilde{m}_{c} \widetilde{m}_{d} }{
      {\cal D}_{ 2u } {\cal D}_{ 1d } } } \left( \sqrt{ \left( 1 - \delta_{u} \right) \left( 1 - 
      \delta_{d} \right) \xi_{ 2 }^u \xi_{ 1 }^d } + \sqrt{ \delta_{u} \delta_{d} \xi_{ 1 }^u \xi_{ 2 }^d }e^{ i \phi_2 }  
      \right) e^{ i \phi_1 },
      \end{array}
    \end{equation}
    \begin{equation}\nonumber
      \begin{array}{l}
    V_{cs}^{ ^{th} } = 
     \sqrt{ \frac{ \widetilde{m}_{u} \widetilde{m}_{d} \xi_{ 2 }^u \xi_{ 2}^d }{ 
      {\cal D}_{ 2u } {\cal D}_{ 2d } } } + \sqrt{ \frac{ \widetilde{m}_{c} \widetilde{m}_{s} }{
      {\cal D}_{ 2u } {\cal D}_{ 2d } } } \left( \sqrt{ \left( 1 - \delta_{u} \right) \left( 1 - 
      \delta_{d} \right) \xi_{ 2 }^u \xi_{ 2 }^d } + \sqrt{ \delta_{u} \delta_{d} \xi_{ 1 }^u \xi_{ 1 }^d }e^{ i \phi_2 }  
      \right) e^{ i \phi_1 }, \\\\
    V_{cb}^{ ^{th} } = 
     - \sqrt{ \frac{ \widetilde{m}_{u} \widetilde{m}_{d} \widetilde{m}_{s} \delta_{d} \xi_{ 2 }^u 
      }{ {\cal D}_{ 2u } {\cal D}_{ 3d } } } + \sqrt{ \frac{ \widetilde{m}_{c} }{ 
      {\cal D}_{ 2u } {\cal D}_{ 3d } } }  \left( \sqrt{ \left( 1 - \delta_{u} \right) \left( 1 
      - \delta_{d} \right) \delta_{d} \xi_{ 2 }^u } - \sqrt{ \delta_{u} \xi_{ 1 }^u \xi_{ 1 }^d \xi_{ 2 }^d 
      }e^{ i \phi_2 } \right) e^{ i \phi_1 } , \\\\
   V_{td}^{ ^{th} } = 
    \sqrt{ \frac{ \widetilde{m}_{u} \widetilde{m}_{c} \widetilde{m}_{s} \delta_{u} \xi_{ 1 }^d }{ 
     {\cal D}_{ 3u } {\cal D}_{ 1d } } } + \sqrt{ \frac{ \widetilde{m}_{d} }{ 
     {\cal D}_{ 3u } {\cal D}_{ 1d } } } \left( \sqrt{ \delta_{u} \left( 1 - \delta_{u} \right) 
     \left( 1 - \delta_{d} \right) \xi_{ 1 }^d } - \sqrt{ \delta_{d} \xi_{ 1 }^u \xi_{ 2 }^u \xi_{ 2 }^d }e^{ i \phi_2 } 
     \right) e^{ i \phi_1 }, \\\\
   V_{ts}^{ ^{th} } = 
    - \sqrt{ \frac{ \widetilde{m}_{u} \widetilde{m}_{c} \widetilde{m}_{d} \delta_{u} \xi_{ 2}^d }{ 
     {\cal D}_{ 3u } {\cal D}_{ 2d } } } + \sqrt{ \frac{ \widetilde{m}_{s} }{ 
     {\cal D}_{ 3u } {\cal D}_{ 2d } } } \left( \sqrt{ \delta_{u} \left( 1 - \delta_{u} \right) 
     \left( 1 - \delta_{d} \right) \xi_{ 2 }^d } - \sqrt{ \delta_{d} \xi_{ 1 }^u \xi_{ 2 }^u \xi_{ 1 }^d }e^{ i \phi_2 } 
     \right) e^{ i \phi_1 },\\\\
   V_{tb}^{ ^{th} } = 
    \sqrt{ \frac{ \widetilde{m}_{u} \widetilde{m}_{c} \widetilde{m}_{d} \widetilde{m}_{s} 
     \delta_{u} \delta_{d} }{ {\cal D}_{ 3u } {\cal D}_{ 3d } } } + \left( \sqrt{ 
     \frac{ \xi_{ 1 }^u \xi_{ 2 }^u \xi_{ 1 }^d \xi_{ 2 }^d }{ {\cal D}_{ 3u } {\cal D}_{ 3d } } } 
     + \sqrt{ \frac{ \delta_{u} \delta_{d} \left( 1 - \delta_{u} \right) \left( 1 - \delta_{d} 
     \right) }{ {\cal D}_{ 3u } D_{ 3d } } }e^{ i \phi_2 } \right) e^{ i \phi_1 }.
  \end{array}
 \end{equation} 
\end{footnotesize} 		
	with
 \begin{footnotesize}
\begin{equation}
  \xi_{1}^{u,d} = 1 - \widetilde{m}_{u,d} - \delta_{u,d} , \quad \xi_{2}^{u,d} = 1 + 
  \widetilde{m}_{c,s} - \delta_{u,d},
  \end{equation}
\end{footnotesize}
  \begin{footnotesize}
\begin{eqnarray}\label{Ds}
  {\cal D}_{ 1(u,d) } = ( 1 - \delta_{u,d} )( \widetilde{m}_{u,d} + \widetilde{m}_{c,s} )( 1 - 
  \widetilde{m}_{u,d} ), \quad  {\cal D}_{ 2(u,d) } = ( 1 - \delta_{u,d} )( \widetilde{m}_{u,d} + \widetilde{m}_{c,s} )( 1 + 
   \widetilde{m}_{c,s} ), \\ \nonumber
   {\cal D}_{3(u,d)} = ( 1 - \delta_{u,d} )( 1 - \widetilde{m}_{u,d} )( 1 + \widetilde{m}_{c,s} ). 
  \end{eqnarray}
\end{footnotesize}
\vspace{-.5cm}

\noindent Here $\tilde m_u=m_u/m_t$, $\tilde m_c=m_u/m_t$, $\tilde m_d=m_d/m_b$ and $\tilde m_s=m_s/m_b$. The parameters $\delta_u$ and $\delta_d$ reflect the mixing between doublet and singlet fermion representations. 
We note that when the phase $\phi_2$ is equal to zero, we recover the model of three $SU(2)_L$ Higgs doublets, two in the doublet representation and one in a singlet symmetric representation of $S_3$. For the general case, when we have the extra Higgs doublet in the anti-symmetric singlet representation of $S_3$, the preferred solution is when $\phi_1=0$, and such that the parameters $\delta_u$ and $\delta_d$ can be treated essentially as free parameters. In this case then, CP violation is achieved by a non zero value of $\phi_2$.
		
	\section{$\chi^2$ analysis for the different models (quark sector)}
\begin{figure}
\centering
\includegraphics[width=6.9cm]{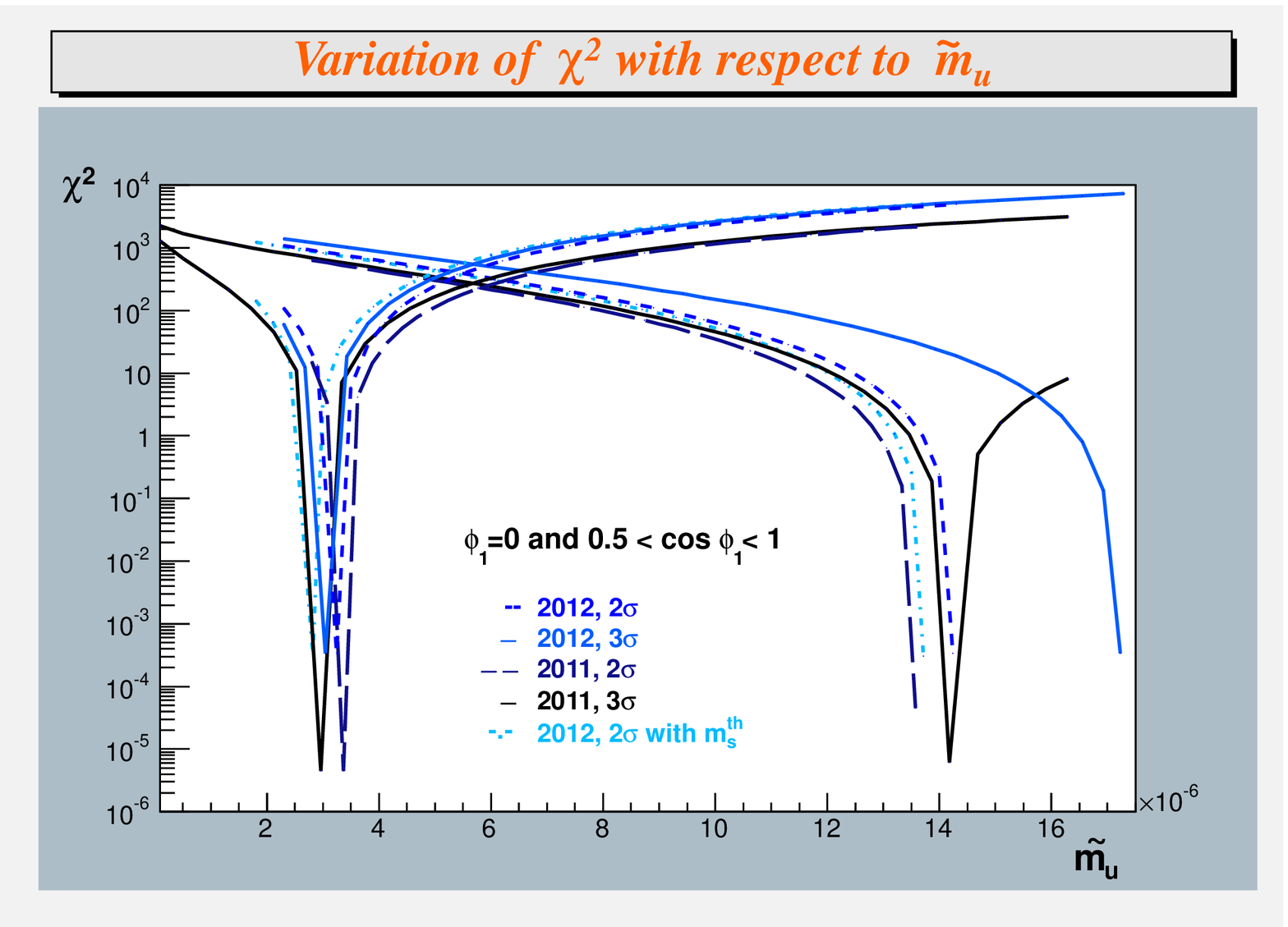}
\includegraphics[width=6.9cm]{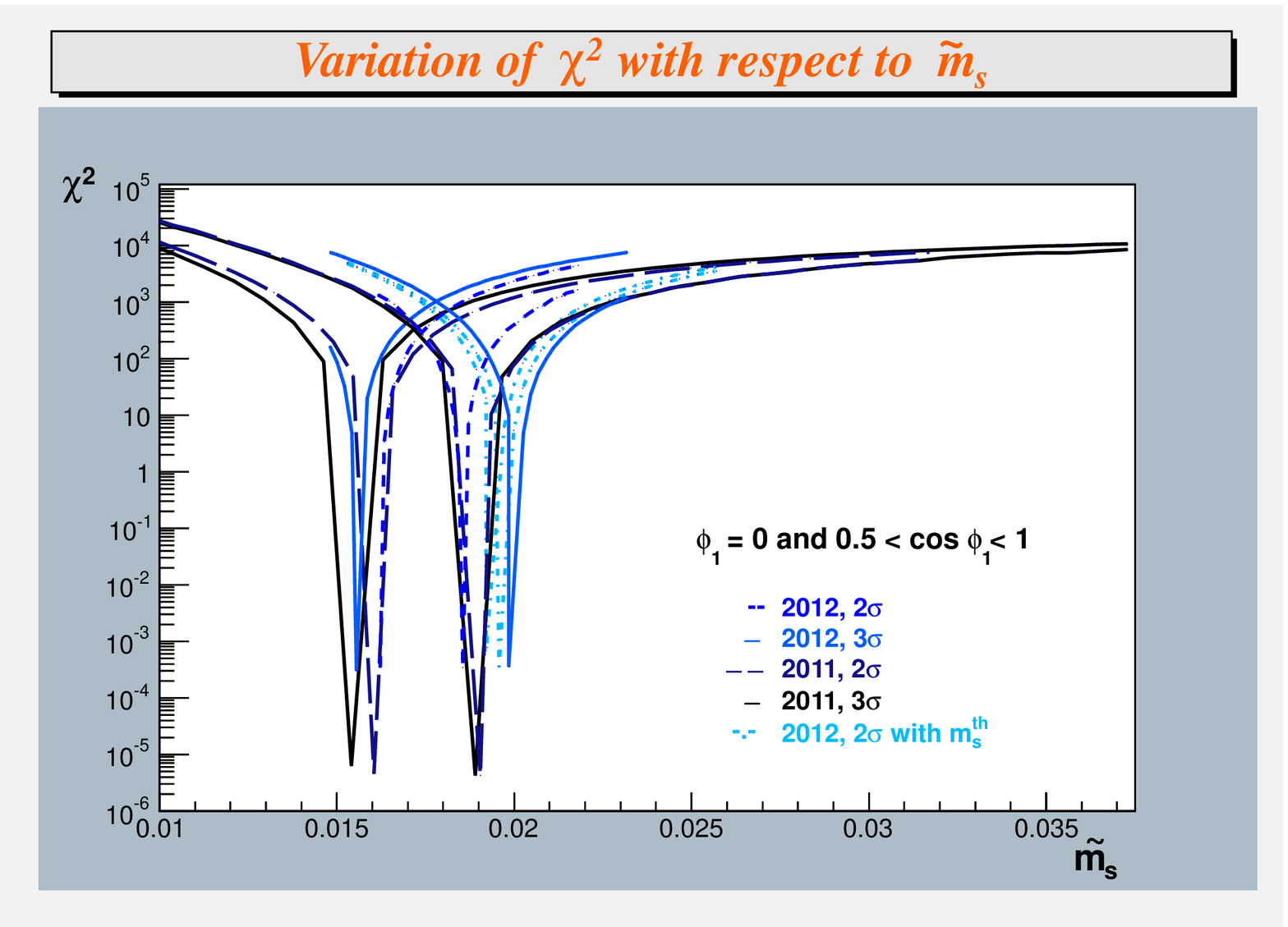}
\caption{\footnotesize{Results of the $\chi^2$ fit as a function of $\widetilde{m_u}$ and  $\widetilde{m_s}$. The values of the mass ratios $\widetilde{m_u}$ and  $\widetilde{m_s}$ reported by the PDG before July 2012 had an uncertainty of about 30\% of their central value. Therefore, they  were  difficult to fit. In July 2012, the up-to-date values of the quark mass ratios were reported by the PDG. The main change was the reduction in the uncertainty of the mass of the strange quark, driven by lattice determinations, which is now about 5\% of its central value. We have also made a fit using only an average of the theoretical determination of $m_s$,  $m_s(2\rm{GeV})=0.101 \pm 0.011$,  in order to asses the impact of the reduction in the uncertainty of $m_s$.  We also show the $\chi^2$ as a function of $\widetilde{m_u}$ because of the size of its uncertainty.  In the figure corresponding to $\widetilde{m_u}$, the plots centered around  $3\times 10^{-6}$ are the result  of the fit where $\phi_1$ is fixed to 0, which corresponds to a version of the model with four Higgs $SU(2)_L$ doublets, two in the doublet representation of $S_3$ and the other two in singlet representations. In this case,  CP violation is a result of the non zero value of $\phi_2$.  In the figure of $\chi^2$ as a function of  $\widetilde{m_s}$, the plots centered around  $1.6\times 10^{-2}$ correspond to the fit where $\phi_1=0$.  The other plots, correspond to the fits of the model with three Higgs fields.}}\label{fig:fitsA}
\end{figure}

For the $\chi^2$ fit we proceed as follows.  We construct the $\chi^2$ function as
\begin{footnotesize}
\begin{eqnarray}
\chi^2=\frac{\left(V_{ud}^{\text{th}}-V_{ud}\right)^2}{\sigma_{V_{ud}}^2}+
\frac{\left(V_{us}^{\text{th}}-V_{us} \right)^2}{\sigma_{V_{us}}^2}+
\frac{\left(V_{ub}^{\text{th}}-V_{ub} \right)^2}{\sigma_{V_{ub}}^2}+
\frac{\left(\mathcal{J}^{\text{th}}_q - \mathcal{J}_q  \right)^2}{\sigma_{{\mathcal{J}_q}}^2},
\end{eqnarray}
\end{footnotesize}

\noindent where the quantities with super-index {{``}}$\text{th}$" are the complete expressions for the CKM elements, as given by the $S_3$ models{,} and those without, the experimental quantities along their  uncertainty, are
\begin{equation}
\label{ex:ckmexpinp_fits}
\begin{array}{llll}
2011: & V_{ud} = 0.97428\pm 0.00015, \hspace{.35cm} &  2012: & V_{ud} =0.97427\pm 0.00015, \\
          & V_{us} = 0.2253\pm 0.007, \hspace{.35cm} &    &  V_{us} =0.2253\pm 0.007, \\
          & V_{ub} = 0.00347\pm 0.00014, \hspace{.35cm} &    &  V_{ub} =0.00351\pm 0.00015,\\
         & J= (2.91\pm 0.155)\times 10^{-5},\hspace{.35cm} &    &J= (2.96\pm 0.18)\times 10^{-5}.
\end{array}
\end{equation}
We present a comparison of fits using  the available data known before July 2012 (labeled as 2011) and the up-to-date data presented by the PDG, given the important change in the uncertainty of the mass of the strange quark.
Since we assume unitarity of the CKM mixing matrix, we need to fit  just to four observables. The theoretical expressions of the parameters of the CKM elements are given in terms of the mass ratios, $\widetilde{m_i}$,  hence the minimization of the defined $\chi^2$ is a function of the {\it parameters}  $\widetilde{m_i}$, $\delta_u$, $\delta_d$, $\cos{\phi_1}$ and $\cos{\phi_2}$. That means, that as a result of the minimization, there is a {\it best fit value}, for each of those quantities, where  $\chi^2$  achieves its minimum.  Of course, the mass ratios  $\widetilde{m}_i$ are not free parameters. The limits that we use correspond to the 3$\sigma$ experimental/lattice regions. The used values for the parameters {$\widetilde{m}_i$}  are given in Table \ref{tab:ratiosmasses}.
\begin{table}[h]
\centering
\begin{tabular}{|c|c|c|}
\hline
\hline
& 2011  &  2012 \\
\hline
$\widetilde {m_u}\left(M_Z\right)$   & $0.0000082\pm 0.0000027$ & $0.0000083\pm 0.0000030$ \\
$\widetilde {m_c}\left(M_Z\right)$   & $0.0036\pm 0.0004$        &  $0.0037\pm 0.00008$ \\
$\widetilde {m_d}\left(M_Z\right)$   & $0.00098 \pm 0.00018$  &  $0.00098\pm 0.00017$ \\
$\widetilde {m_s}\left(M_Z\right)$   & $0.0205 \pm 0.0056$       &  $0.0190\pm 0.0014$ \\
\hline
\end{tabular}
\caption{Comparison of the values of the mass ratios, at $ M_Z$, reported up to 2011 and the newest values reported by the PDG.
}
\label{tab:ratiosmasses}
\end{table}
We used MINUIT and varied all the parameters {{$\widetilde{m}_i$}},  within the $3\sigma$ range given in Table \ref{tab:ratiosmasses}, and $\delta_u$,  $\delta_d$, as really free parameters. The condition that $\phi_1=0$ is imposed by consistency in finding appropriate solutions for the case of the four Higgs fields that we are considering here. In this case, CP violation arises as a result of having $\phi_2\neq 0$. 
 Minimization with MINUIT relies on starting the process  with a seed, for all parameters, close enough to the minimum. Therefore, if the range of variation is large, it is difficult to find a best fit point. On the other hand, to check for global minima, one should remove the limits of the "free" parameters. Unfortunately, if we perform the fit leaving completely free the parameters  $\widetilde{m_i}$, the quality of the fit decreases, and most importantly,  it would turn out that $\widetilde{m_u}$ would be  of order $10^{-3}$. In Figure \ref{fig:fitsA} we present the result of the fit for the case that we considered here and we compare it to the case of the model with three Higgs fields.  For the case of four Higgs doublets, the values found for $\delta_u$, $\delta_d$ and $\phi_2$ are respectively $7.9\times 10^{-2}$, $3.9\times 10^{-2} $ and $0.469$ radians. For the case of three Higgs doublets, the values  found for $\delta_u$, $\delta_d$ and $\phi_1$ are respectively $2.5\times 10^{-2}$, $4.09\times 10^{-2} $ and $1.047$ radians.

\section*{Conclusions}
			
		 In order to explore the Higgs sector of $S_3$ family models, we studied a model containing four Higgs $SU(2)_L$ doublets belonging to the $S_3$  representation  $4 = {\bf{2}} \oplus {\bf{1}}_{S} \oplus {\bf{1}}_{A}$. We were able to identify the conditions under which the {4-zero Fritzsch-like texture} mass matrices are obtained. This allowed us to reduce the number of parameters needed to describe the generic model with four Higgs $SU(2)_L$ doublets and to find an exact parameterization of the mass matrices in terms of their mass eigenvalues. We provided also exact formulas for mixing angles of the CKM matrix in terms of quark mass ratios. This line of work was developed  in \cite{Barranco:2010we}, without a particular model in mind, but it was clear the usefulness in classifying mass matrix patterns according to their transformation to a {4-zero Fritzsch-like texture}. The reduction to this form, then allowed us to make a direct comparison among different models and to perform a $\chi^2$ analysis to the CKM elements.  A particular case of the general model presented, is a case for which only three Higgs $SU(2)_L$ doublets are allowed  and for which we effectively reduced one free parameter. Finally, we have then proceeded to perform a $\chi^2$ analysis to the CKM elements, both to the general model, $S_3$-SM, and to the restricted one. In both cases, we have obtained  impressive values of the minimum of the $\chi^2$ function, $3.3 \times 10^{-4}$ and $3.4 \times 10^{-4}$, respectively. 
			
	\section*{Acknowledgements}
	U. J. S. S. thanks the Instituto de F\'isica-UNAM for supporting his attendance to this interesting and stimulating workshop. We acknowledge support from the Mexican grants: PAPIIT IN113712 and CONACyT-132059.
	
\section*{References}
\bibliography{S3_Aug2012}

\end{document}